%
%
\input harvmac.tex
%
\def\cn{{\cal N}}
\def\IR{\relax{\rm I\kern-.18em R}}
\def\IZ{\relax\ifmmode\hbox{Z\kern-.4em Z}\else{Z\kern-.4em Z}\fi}

%
%
\def\np#1#2#3{{\it Nucl. Phys.} {\bf B#1} (#2) #3}
\def\pl#1#2#3{{\it Phys. Lett.} {\bf #1B} (#2) #3}
\def\plb#1#2#3{{\it Phys. Lett.} {\bf #1B} (#2) #3}
\def\prl#1#2#3{{\it Phys. Rev. Lett.} {\bf #1} (#2) #3}

\def\prep#1#2#3{{\it Phys. Rept.} {\bf #1} (#2) #3}
\def\prd#1#2#3{{\it Phys. Rev.} {\bf D#1} (#2) #3}
\def\atmp#1#2#3{{\it Adv. Theor. Math. Phys.} {\bf #1} (#2) #3}
\def\jhep#1#2#3{{\it J. High Energy Phys.} {\bf #1} (#2) #3}
\def\cqg#1#2#3{{\it Class. Quant. Grav.} {\bf #1} (#2) #3}
%
%

\lref\witm{ E. Witten, 
``Solutions of four-dimensional field theories via M-theory,''
hep-th/9703166, 
\np{500}{1997}{3}.}

\lref\juan{J. M. Maldacena, ``The large $N$ limit of superconformal field
theories and supergravity,'' hep-th/9711200, \atmp{2}{1998}{231}.}

\lref\gkp{S. S. Gubser, I. R. Klebanov and A. M. Polyakov, 
``Gauge theory correlators
from non-critical string theory,'' hep-th/9802109,
\plb{428}{1998}{105}.}

\lref\wittenads{E. Witten, ``Anti-de-Sitter space and holography,''
hep-th/9802150, \atmp{2}{1998}{253}.}

\lref\magoo{O. Aharony, S. S. Gubser, J. M. Maldacena, H. Ooguri and
Y. Oz, ``Large $N$ field theories, string theory and gravity,''
hep-th/9905111, \prep{323}{2000}{183}.}

\lref\polstr{J. Polchinski and M. J. Strassler, ``The string dual of a
confining four-dimensional gauge theory,'' hep-th/0003136.}

\lref\donwit{R.~Donagi and E.~Witten, ``Supersymmetric Yang-Mills
theory and integrable systems,'' hep-th/9510101,
\np{460}{1996}{299}.}

\lref\swone{N.~Seiberg and E.~Witten,
``Electric-magnetic duality, monopole condensation, and confinement
in $\cn=2$ supersymmetric Yang-Mills theory,'' hep-th/9407087,
\np{426}{1994}{19}.}

\lref\swtwo{N.~Seiberg and E.~Witten,
``Monopoles, duality and chiral symmetry breaking in $\cn=2$ 
supersymmetric QCD,'' hep-th/9408099, \np{431}{1994}{484}.}

\lref\newpol{J.~Polchinski,
``$\cn=2$ gauge-gravity duals,'' hep-th/0011193, contributed to
Strings 2000, Ann Arbor, Michigan, submitted to
Int. J. Mod. Phys. {\bf A}.}

\lref\ks{
S.~Kachru and E.~Silverstein,
``4d conformal theories and strings on orbifolds,'' hep-th/9802183,
\prl{80}{1998}{4855}.}

\lref\bdflmp{
M.~Bertolini, P.~Di Vecchia, M.~Frau, A.~Lerda, R.~Marotta and I.~Pesando,
``Fractional D-branes and their gauge duals,''
hep-th/0011077.}

\lref\klenek{
I.~R.~Klebanov and N.~A.~Nekrasov,
``Gravity duals of fractional branes and logarithmic RG flow,''
hep-th/9911096, \np{574}{2000}{263}.}

\lref\klestr{
I.~R.~Klebanov and M.~J.~Strassler,
``Supergravity and a confining gauge theory: Duality cascades and 
$\chi$SB-resolution of naked singularities,'' hep-th/0007191,
\jhep{0008}{2000}{052}.}

\lref\leistr{
R.~G.~Leigh and M.~J.~Strassler,
``Exactly marginal operators and duality in four-dimensional $\cn=1$ 
supersymmetric gauge theory,'' hep-th/9503121, \np{447}{1995}{95}.}

\lref\kletse{
I.~R.~Klebanov and A.~A.~Tseytlin,
``Gravity duals of supersymmetric $SU(N) \times SU(N+M)$ gauge theories,''
hep-th/0002159, \np{578}{2000}{123}.}

\lref\klegub{
S.~S.~Gubser and I.~R.~Klebanov,
``Baryons and domain walls in an $\cn = 1$ superconformal gauge theory,''
hep-th/9808075, \prd{58}{1998}{125025}.}

\lref\klewit{
I.~R.~Klebanov and E.~Witten,
``Superconformal field theory on threebranes at a Calabi-Yau  singularity,''
hep-th/9807080, \np{536}{1998}{199}.}

\lref\gubser{
S.~S.~Gubser,
``Einstein manifolds and conformal field theories,''
hep-th/9807164, \prd{59}{1999}{025006}.}

\lref\seiberg{
N.~Seiberg,
``Electric-magnetic duality in supersymmetric non-Abelian gauge theories,''
hep-th/9411149, \np{435}{1995}{129}.}

\lref\mylst{
O.~Aharony,
``A brief review of 'little string theories','' hep-th/9911147,
contributed to Strings '99, \cqg{17}{2000}{929}.}

\lref\abks{O. Aharony, M. Berkooz, D. Kutasov and N. Seiberg, ``Linear
Dilatons, NS5-branes and Holography,'' hep-th/9808149, 
\jhep{9810}{1998}{004}.}

\lref\krasnitz{
M.~Krasnitz,
``A two point function in a cascading $\cn = 1$ 
gauge theory from  supergravity,''
hep-th/0011179.}

\lref\buchel{
A.~Buchel,
``Finite temperature resolution of the Klebanov-Tseytlin singularity,''
hep-th/0011146.}

\lref\clp{M.~Cvetic, H.~Lu and C.~N.~Pope,
``Brane resolution through transgression,''
hep-th/0011023.}

\lref\cglp{M.~Cvetic, G.~W.~Gibbons, H.~Lu and C.~N.~Pope,
``Ricci-flat metrics, harmonic forms and brane resolutions,''
hep-th/0012011.}

\lref\jpp{C.~V.~Johnson, A.~W.~Peet and J.~Polchinski,
``Gauge theory and the excision of repulson singularities,''
hep-th/9911161, \prd{61}{2000}{086001}.}

\lref\ejp{N.~Evans, C.~V.~Johnson and M.~Petrini,
``The enhancon and $\cn = 2$ gauge theory/gravity RG flows,''
hep-th/0008081, \jhep{0010}{2000}{022}.}

\lref\bpp{A.~Buchel, A.~W.~Peet and J.~Polchinski,
``Gauge dual and noncommutative extension of an $\cn = 2$ 
supergravity  solution,'' hep-th/0008076.}

\lref\cliff{
C.~V.~Johnson,
``The enhancon, multimonopoles and fuzzy geometry,''
hep-th/0011008.}

\lref\paul{P. S. Aspinwall, ``Enhanced Gauge Symmetries and K3
Surfaces,'' hep-th/9507012, \pl{357}{1995}{329}.}

\lref\arvind{
A. Rajaraman, ``Supergravity duals for $\cn=2$ gauge theories,''
hep-th/0011279.}

\lref\klewitt{
I.~R.~Klebanov and E.~Witten,
``AdS/CFT correspondence and symmetry breaking,'' hep-th/9905104,
\np{556}{1999}{89}.}

\lref\malnun{
J.~M.~Maldacena and C.~Nunez,
``Towards the large $N$ limit of pure $\cn = 1$ 
super Yang Mills,'' hep-th/0008001.}

\lref\oldme{
O. Aharony, ``Remarks on nonabelian duality in $\cn=1$ supersymmetric
gauge theories,'' hep-th/9502013, \pl{351}{1995}{220}.}

\lref\aps{
P.~C.~Argyres, M.~Ronen Plesser and N.~Seiberg,
``The moduli space of $\cn=2$ SUSY QCD and duality in $\cn=1$ SUSY QCD,''
hep-th/9603042, \np{471}{1996}{159}.}

\lref\cacher{
E.~Caceres and R.~Hernandez,
``Glueball masses for the deformed conifold theory,''
hep-th/0011204.}

\lref\fgpw{
D.~Z.~Freedman, S.~S.~Gubser, K.~Pilch and N.~P.~Warner,
``Continuous distributions of D3-branes and gauged supergravity,''
hep-th/9906194, \jhep{0007}{2000}{038}.}

\lref\leoark{L.~A.~Pando Zayas and A.~A.~Tseytlin,
``3-branes on resolved conifold,'' hep-th/0010088, \jhep{0011}{2000}{028}.}

\lref\newgub{S.~S.~Gubser,
``Supersymmetry and F-theory realization of the deformed conifold with  
three-form flux,'' hep-th/0010010.}

\lref\enrs{I.~P.~Ennes, S.~G.~Naculich, H.~Rhedin and H.~J.~Schnitzer,
``One-instanton predictions for non-hyperelliptic curves derived from  
M-theory,'' hep-th/9806144, \np{536}{1998}{245}.}

\lref\enrstwo{I.~P.~Ennes, S.~G.~Naculich, H.~Rhedin and H.~J.~Schnitzer,
``Tests of M-theory from $\cn = 2$ Seiberg-Witten theory,''
hep-th/9911022.}

\Title{\vbox{\baselineskip12pt
\hbox{hep-th/0101013}\hbox{WIS/27/00-DEC-DPP}}}
{\vbox{
{\centerline {A Note on the Holographic Interpretation of}}
\vskip .1in
{\centerline {String Theory Backgrounds with Varying Flux}}
  }}
\centerline{Ofer Aharony\foot{e-mail : \tt ofer.aharony@weizmann.ac.il.}}
\vskip.1in 
\centerline{ \it Department of Particle Physics }
\centerline{ \it The Weizmann Institute of Science }
\centerline{ \it Rehovot 76100, Israel }

\vskip .2in

\centerline{\bf Abstract}

\vskip .1in

We discuss the field theory interpretation (via holographic duality)
of some recently-discovered string theory solutions with varying flux,
focusing on four dimensional theories with $\cn=2$ supersymmetry and
with $\cn=1$ supersymmetry which arise as the near-horizon limits of
``fractional D3-branes''. We argue that in the $\cn=2$ case the best
interpretation of the varying flux in field theory is via a Higgs
mechanism reducing the rank of the gauge group, and that there is no
need to invoke a duality to explain the varying flux in this case. We
discuss why a similar interpretation does not seem to apply to the
$\cn=1$ case of Klebanov and Strassler, which was interpreted as a
``duality cascade''. However, we suggest that it might apply to
different vacua of the same theory, such as the one constructed by
Pando Zayas and Tseytlin.

\Date{12/00}

\newsec{Introduction and Summary}

It is very interesting to generalize the AdS/CFT correspondence
\refs{\juan,\gkp,\wittenads} (see \magoo\ for a review) to
theories without conformal invariance. On the field theory side
non-conformal theories can have richer dynamics, and are more
directly related to experiment. On the string theory side, such a
generalization enables a holographic interpretation of additional
backgrounds beyond anti-de Sitter spaces.

The simplest way to obtain such a generalization is to start from a
conformal field theory with a known AdS dual, and then deform it by
relevant operators, and/or go to its moduli space (if it has one),
thus breaking conformal invariance explicitly or spontaneously. In
these cases the string theory background is still asymptotically AdS,
with known behaviors near the boundary for various fields,
corresponding to the deformations and vacuum expectation values (VEVs)
we turn on in the field theory. Various examples of this type have
been studied, one of the more interesting ones (which is related in
various limits to $\cn=1$ SYM and to pure Yang-Mills theory) being the
mass deformation of the $\cn=4$ SYM theory, whose string theory dual
was found in \polstr. 

An alternative way to get non-conformal theories is to study
gravitational backgrounds with a different asymptotic behavior. For
example, backgrounds which asymptote to linear dilaton backgrounds
were argued in \abks\ to be dual to ``little string theories''.  In
fact, this seems to be the only general (Lorentz-invariant) example
which is well-understood. Different asymptotic behaviors would
correspond to theories whose UV behavior is neither that of a local
field theory nor that of a ``little string theory'', and we do not
know of any other possible behaviors for Lorentz-invariant theories.

An interesting example of a background with a different asymptotic
behavior was found in \klestr\ (following \refs{\klenek,\kletse}) by
examining the near-horizon limit of $N$ D3-branes and $M$ fractional
D3-branes at a conifold point. The background of \klestr\ is
completely non-singular, but its asymptotic behavior is different from
all previously known examples. In particular, the 5-form flux on the
compact 5-cycle grows with the radial direction, and diverges at
infinity, suggesting that the background may correspond to the large
$N$ limit of some field theory (unlike the usual examples which are
dual to field theories of finite $N$, though $N$ may be large). The
2-point functions in this background \krasnitz\ and its finite
temperature behavior \buchel\ seem to agree with this
interpretation, and it would be interesting to make this
interpretation more precise. Masses of some of the scalar particles in
the theory corresponding to this background were recently computed in
\cacher.

Several other examples with the same type of behavior have since been
discovered. In \refs{\clp,\cglp}, various backgrounds in which the
flux varies were discussed. Some of these backgrounds are
asymptotically AdS, so they have a simple interpretation as a
deformation of a conformal theory (and/or a point on its moduli
space), but in the other backgrounds the flux diverges at infinity
so they are of the same type as the background of \klestr\ (though
generically the divergence of the flux in these backgrounds is much
worse than the logarithmic divergence of
\klestr, so it is not clear if they are really of the same class). In
\refs{\newpol,\bdflmp} the near-horizon limit of fractional branes at
an $\IR^4/\IZ_2$ orbifold point was studied, and found to lead to a
similar behavior of the flux\foot{The same configuration was studied
with different boundary conditions in \arvind.}. The authors of
\leoark\ constructed another solution that has the same asymptotic
behavior as the solution of \klestr, but a different structure in the
interior of space, which contains a repulson-like singularity. If this
singularity can be resolved this background should correspond to
another vacuum of the same field theory as in \klestr. Clearly, many other
solutions with the same behavior can also be constructed by similar
methods.

It seems natural to associate the changing flux with a decrease in the
number of degrees of freedom of the field theory as one decreases the
energy. The authors of \klestr\ suggested that in the case with $\cn=1$
supersymmetry
which they discussed\foot{The background of \klestr\ was shown to be
supersymmetric in \refs{\newgub,\cglp}.}, this decrease comes
from a series of duality transformations reducing the size of the
gauge group. In this paper we would like to suggest that in other cases,
such as the $\cn=2$ case of \refs{\newpol,\bdflmp}, there is actually
a different, simpler interpretation of this decrease.
The theories involved in the
backgrounds discussed here all have large moduli spaces, corresponding
to putting the branes at other positions rather than the origin
(before taking the near-horizon limit).  We suggest that in the case
of \refs{\newpol,\bdflmp}, and perhaps also in some of the other cases,
the gauge theory dual to the known string theory backgrounds 
is at a point on its moduli space such that the distribution of
the branes exactly mirrors the source for the corresponding field in
the string theory; namely, that in the field theory there ``really
are'' D3-branes where there is a source for the 5-form flux in the
string theory (with obvious generalizations to theories in other
dimensions). At this position in the moduli space the gauge group is 
spontaneously broken by the
Higgs mechanism, leading naturally to a reduction in the size of the
group. Of course, the gauge theory could also be at a point with
zero VEVs (at least classically), but we claim that this point does
not correspond to the known string theory duals.
This seems to be the most naive interpretation of the varying
flux, and we will see that in the $\cn=2$ theories of \refs{\newpol,\bdflmp}
it seems to be implied by the VEVs we
compute in such backgrounds.

In the case with $\cn=2$ supersymmetry discussed in \refs{\newpol,\bdflmp}
we will be able to test this
conjecture in various ways, since in this case we know how to define
the theory as a limit of well-defined field theories, and since we can
use the information from the effective action of wrapped D5-branes in
the corresponding background. This case will be described in detail in
section 2. We do not perform any new non-trivial computations, but
just review how the existing results are consistent with the
``Higgsing interpretation''. 

In section 3 we will discuss the case of \klestr\ with $\cn=1$
supersymmetry. In this case it seems that the same interpretation does
not apply, so the decreasing flux is probably best thought of as
coming from a ``duality cascade'' as described in \klestr.  A
different vacuum of the same theory was constructed in \leoark, and if
the singularity there can be resolved, we suggest that in this vacuum
the decrease in the size of the gauge group could perhaps be
interpreted as Higgsing, like in the $\cn=2$ case. For the cases with
$\cn=1$ supersymmetry the evidence for the identification of the
position in the moduli space 
and for the interpretation of the varying flux is much weaker, and
it would be nice to understand them better.
 
Similar methods can be used to analyze other cases, such as the
backgrounds constructed in \refs{\clp,\cglp}, to see whether they are
well-described by a Higgsing interpretation, or if dualities are
needed to explain the decrease in the flux.  Of course, in the cases
of $d\neq 4$ which are not dual to gauge theories the ``Higgsing
interpretation'' does not literally involve a Higgs mechanism, but the
relevant physics (in terms of where the theory sits in the moduli space)
is completely analogous. We will not discuss these other cases here.

\newsec{Theories Related to Branes on an $\IR^4/\IZ_2$ Singularity}

\subsec{The Conformal Case}

We will begin by discussing the theory of $N$ D3-branes on an
$\IR^4/\IZ_2$ singularity in type IIB string theory. The low-energy
theory on these branes is the $U(N)\times U(N)$ $\cn=2$ gauge theory
with 2 bifundamental hypermultiplets. The sum of the inverse gauge
couplings $\tau_i = {4\pi i \over {(g_{YM}^2)_i}} + {\theta_i \over 
{2\pi}}, i=1,2$, 
is related to the string coupling $\tau = {i \over g_s} + {\chi \over 2\pi}$,
while their difference is related to the integrals of the 2-form
fields over the 2-cycle which vanishes at the orbifold
singularity. The appearance of two gauge groups may be interpreted as arising
from fractional D3-branes, or D5-branes (and anti-D5-branes)
wrapped on the vanishing
2-cycle, as described in detail (for instance) in \newpol. The beta
functions of both $SU(N)$ factors vanish, while the diagonal $U(1)$ is
free and decoupled, and the off-diagonal $U(1)$ is free (and becomes a
global symmetry) in the IR.

The classical moduli space of this field theory agrees precisely with the
configurations of (possibly fractional) branes moving on the
orbifold. There is a ``Higgs branch'' where the hypermultiplets obtain
VEVs, in which case a combination of the adjoint fields becomes
massive (while the other combination can also acquire a VEV). The
moduli space of this branch is of the form $(\IR^4/\IZ_2 \times
\IR^2)^N / S_N$, corresponding to the motion of D3-branes in the
background. On this branch there are no quantum corrections to the metric on
the moduli space, both on the field theory side and on the string
theory side.  There is also a ``Coulomb branch'' in which the
hypermultiplet VEVs vanish but the two adjoint fields (in the $\cn=2$
vector multiplets), $\varphi_1$ and $\varphi_2$, acquire VEVs, whose
eigenvalues may be interpreted as positions of fractional D3-branes
(or wrapped D5-branes). This branch is of the form $(\IR^2)^N / S_N
\times (\IR^2)^N / S_N$. There are also mixed branches which
correspond to configurations with both D3-branes and wrapped D5-branes
in an obvious way.

The near-horizon limit of this brane configuration is related by the
AdS/CFT correspondence \refs{\juan,\gkp,\wittenads,\ks} 
to the non-trivial part of the 
low-energy field theory on the branes, which is the
$SU(N)\times SU(N)$ $\cn=2$ supersymmetric gauge theory with 2 bifundamental
hypermultiplets. The geometry of this near-horizon limit is \ks\
$AdS_5 \times S^5/\IZ_2$, where the $\IZ_2$ action leaves fixed an
$S^1$ inside the $S^5$. The global symmetry of the field theory is
$SU(2)\times SU(2)\times U(1)_R\times U(1)_B$. The first three factors
are realized as $SO(4)\times SO(2)$ isometries of $S^5/\IZ_2$, while
the fourth corresponds to a gauge field from the twisted sector, as
described below\foot{This gauge field 
can be thought of as the 4-form of type IIB
supergravity integrated over the 2-cycle which vanishes at the
orbifold and over the fixed $S^1$.}.

The field theory has a global $\IZ_2$ symmetry exchanging the two
gauge groups (which is broken when they have different gauge
couplings), and this may be identified with the $\IZ_2$
symmetry of the orbifold, under which twisted sector states are
charged. Thus, operators in the field theory which are symmetric under
this $\IZ_2$ are identified with untwisted sector states which
propagate on the full $AdS_5\times S^5/\IZ_2$, while anti-symmetric
operators are identified with twisted sector states which propagate
only on the fixed line $AdS_5\times S^1$. Generically the light states
living on the fixed line are just a six dimensional tensor
multiplet. However, when the two 2-form fields integrated over the
vanishing 2-cycle vanish, there are additional light states coming
from D3-branes wrapped on the vanishing 2-cycle, which give rise to
tensionless strings; in the limit $g_s \to 0$ these strings give the
$A_1$ $\cn=(2,0)$ ``little string theory'' (see \mylst\ for a review
with references on ``little string theories''). In the low-energy theory
on $AdS_5$ these strings give rise to an $SU(2)$ gauge symmetry;
presumably this can be identified with an $SU(2)$ global symmetry of
the corresponding field theory which arises when one of the couplings
becomes infinite\foot{A $U(1)$ subgroup of this $SU(2)$ is the
$U(1)_B$ global
symmetry which arises from the off-diagonal $U(1)$ in $U(N)\times U(N)$.}.

There is no general method to construct the moduli space of a field
theory from its holographic dual. One has to construct separate
solutions for each possible value of the VEVs (which is a different
superselection sector in the field theory) and verify that they have
the same UV behavior (so they correspond to the same field theory),
with appropriate boundary values for the string theory modes that
correspond to the fields which acquire expectation values. In the case
of the $SU(N)\times SU(N)$ $\cn=2$ theory described above, it is not
hard to construct solutions corresponding to any configuration in the
moduli space. 

First, let us examine configurations on the Higgs branch,
corresponding to moving around the D3-branes. The description of these
configurations is the same as that of the moduli space of the $\cn=4$
SYM theory. For the $\cn=4$ theory, the dual background is
\eqn\ads{ds^2 = Z^{-1/2} \eta_{\mu \nu} dx^{\mu} dx^{\nu} + Z^{1/2}
dx^m dx^m}
where $\mu,\nu=0,1,2,3$, $m,n=4,5,6,7,8,9$, and $Z = 4\pi g_s N
\alpha'^2 / r^4$ with $r^2 = x^m x^m$. There is also a 5-form field
proportional to the derivative of $Z$, which we
will not write explicitly. To describe configurations on the Coulomb
branch of this theory, one simply replaces $Z$ by a more general
harmonic function 
\eqn\harmonic{Z = 4\pi g_s \alpha'^2 \sum_{j=1}^N {1\over
{|\vec{x}-\vec{x}_j|^4}},} 
where the $\vec{x}_j$ are related to the field
theory VEVs (the eigenvalues $\vec{\Phi}^j$ of the six matrices $\vec{\Phi}$)
by $\vec{\Phi}^j = \vec{x}_j / 2\pi \alpha'$. This is
because the eigenvalue VEVs behave like D3-branes, which are sources for $Z$
(appearing in the metric and in the 5-form).
This identification can be tested by computing the VEVs of operators
such as $\tr(\vec{\Phi}^n)$ in this background (see, for example,
\klewitt).
For general configurations
on the Coulomb branch, this supergravity background is singular
(corresponding to the fact that the field theory is free in the
IR). Configurations where all branes are in big clumps have a good
description in supergravity. Alternatively, configurations with large
$N$ but with only a small number of non-zero VEVs may be described in
terms of D3-brane probes propagating in the background generated by
the other branes, since their back-reaction is negligible in the large
$N$ limit. For example, the configuration where only one VEV $\vec{\Phi}^1$
is non-zero is dual to the string theory background with $Z = 4\pi g_s
(N-1) \alpha'^2 / r^4$, with a single D3-brane at the position $\vec{x} =
2\pi \alpha' \vec{\Phi}^1$. 

The moduli space of the Higgs branch of
the orbifold theory may
be simply derived by orbifolding these backgrounds by $x^m \to -x^m,
m=6,7,8,9$. The harmonic functions are the same, except that we have
to add also an image for every D3-brane \klewitt, so that
\eqn\newharmonic{Z = 4\pi g_s \alpha'^2 \sum_{j=1}^N \left( {1\over
|\vec{x}-\vec{x}_j|^4} + {1\over |\vec{x}-\vec{\tilde x}_j|^4} \right), }
where the reflection takes $x \to \tilde{x}$, and the total flux (now
defined by integrating the 5-form over $S^5/\IZ_2$) is still $N$.

To describe the Coulomb branch, corresponding to wrapped
D5-branes, we have to take into account the fact that these are
sources for the 3-form field strengths of type IIB string theory, which are
usually written in the combination $G_3 = F_3 - \tau H_3$. The wrapped
D5-branes have to sit at the fixed point of the orbifold, so they
only move in $x_4$ and $x_5$, and it will be convenient to label their
position by the complex variable $z = x_4 + ix_5$. It turns
out \refs{\newpol,\bdflmp}
that for the moduli space configurations it is enough to turn on
the components of the 2-form along the vanishing 2-cycle whose volume
form we denote by $\omega_2$, so that $G_3 = d(\theta
\omega_2)$. $\theta$ may be written as $\theta_C - \tau \theta_B$,
where $\theta_{B,C}$ are the integrals of the corresponding 2-forms on
the vanishing 2-cycles, normalized to have periodicity $2\pi$. One
then finds \refs{\newpol,\bdflmp}
a simple Laplace equation for $\theta$, with positive-sign
sources at the positions $z_i$ of wrapped D5-branes and negative-sign
sources at the positions $\tilde{z}_i$ of wrapped anti-D5-branes, and
the solution is 
\eqn\thetasol{\theta = 2i[\sum_{j=1}^N \ln(z-\tilde{z}_j) -
\sum_{j=1}^N \ln(z-z_j)] - \theta_B^0 \tau,}
where $\theta_B^0$
denotes the value of $\theta_B$ in the theory at the origin of moduli
space (which in the orbifold theory is $\theta_B^0=\pi$ \paul)\foot{We
could also have a non-zero $\theta_C^0$, but this will not play an
important role so we set $\theta_C^0=0$.}. 
The equation for $Z$ is now more
complicated than the Laplace equation, since the varying 2-form fields
result in an additional source for the metric and for the 5-form field,
$dF_5 = -F_3\wedge H_3$ (recall that $H_3 = dB_2^{NS-NS}$, $F_3 =
dC_2$ and $F_5 = dC_4 - C_2 \wedge H_3$, where $C_k$ is the $k$-form
RR field). This is in addition to the source coming from the direct
coupling of the wrapped D5-branes to the 5-form field.
The equation and
its solution for coinciding wrapped D5-branes may be found in \bdflmp.
We will not need its explicit form here. 

The 5-form flux varies in the
solution because of the contribution of the 3-form fields to this
flux, in addition to the naive variation coming from the D3-brane
charge carried by the wrapped 5-branes. 
To get a well-defined solution one has to be careful that
this flux does not become negative (of course, for supergravity to be
valid we actually require that this flux, as well as the flux
multiplied by the string coupling, are much greater than one). 
Generally this puts some limit on
the possible positions of the wrapped D5-branes. In the field theory
the $z_j$ correspond to the eigenvalues of the VEV of the first 
$SU(N)$-adjoint field
$\varphi_1$, and the $\tilde{z}_j$ are the eigenvalues of the VEV of
the second $SU(N)$-adjoint field $\varphi_2$. The moduli space
metric of
the field theory is now corrected (if $z_j \neq \tilde{z}_j$), both at
1-loop and from instanton effects, and the classical
identification of the $z_j$ with eigenvalues of the adjoint fields is
no longer exact. The Seiberg-Witten curve \refs{\swone,\swtwo} for
this field theory was found in \witm, but (as far as we know) there
is no explicit form for this curve in terms of the gauge-invariant
operators in the field theory (which are classically related to the
$z_j$)\foot{The curve was found using the brane construction of these
field theories. The other theories we discuss here also have brane
constructions, but it is not clear to us how to use them to study the
issues we discuss here. The one-loop and one-instanton corrections to
the classical curve have been expressed in terms of gauge-invariant
observables in the case of a single bifundamental hypermultiplet in
\enrs, and presumably this can be generalized also to the conformal
case of two bifundamental hypermultiplets, perhaps along the lines of
\enrstwo, but this has not yet been done as far as we know.}. 
Generally, the quantum corrections in the SW curve limit the
possible values of the $z_j$. For example, in the pure $SU(2)$ theory,
the configuration with a vanishing (generalized) VEV $a$ is not on the
quantum moduli space. We expect that in the $SU(N)\times SU(N)$ theory
the quantum corrections arising from the SW curve will result in a
positive 5-form flux for all the
configurations on the quantum moduli space,
at least whenever the supergravity approximation is
valid. It would be very interesting to verify this prediction of the
AdS/CFT correspondence directly.

\subsec{The Non-Conformal Case and its Suggested Interpretation}

Instead of looking at just $N$ D3-branes at the orbifold point, we
could look at $N$ D3-branes plus $M$ wrapped D5-branes (if we have
also wrapped anti-D5-branes we can replace a wrapped D5-brane and a wrapped
anti-D5-brane by a D3-brane, 
so this is the most general case). The same arguments
imply that the low-energy field theory in this case is the $SU(N)\times
SU(N+M)$ $\cn=2$ gauge theory with 2 bifundamental
hypermultiplets. However, for any $M > 0$ this field theory is not
asymptotically free (the beta function of one of the gauge factors is
positive and of the other is negative), 
so there is no limit where it can be decoupled
from string theory and result in some well-defined dual background.

Nevertheless, people have boldly attempted to find duals for this
theory, first perturbatively in $M/N$ in \klenek\ and later exactly
(in the supergravity approximation) in \refs{\newpol,\bdflmp}. The
solutions they found have 3-form fields of the form described above, with
$\theta=-2iM\ln(z/z_0)$ 
for some (arbitrary) cutoff $z_0$. In these
solutions there is an IR singularity at $|z|=r_e$, similar to the famous 
enhan\c con \refs{\jpp,\bpp,\ejp,\cliff}, 
which is presumably resolved in a similar
way. There is also a problem in the UV (large $|z|$), which is that
the 5-form flux grows logarithmically there, and is not bounded by
$N$, so it is not really possible to interpret these theories as dual
to a particular $SU(N)\times SU(N+M)$ gauge theory. This problem is
related to the fact that this gauge theory is not asymptotically free
so it is not well-defined.

Using the results of the previous subsection, we can identify how
these solutions arise from solutions of string theory whose field
theory interpretation is known. If
we look in the moduli space of the $SU(N+M)\times SU(N+M)$ theory
at a configuration where $M$ of the $\tilde{z}_j$ are equal to $(-z_0)$
(and all the other $z_j$'s and $\tilde{z}_j$'s vanish), we have 
\eqn\thetaone{\theta =
-2iM\ln(z/(z+z_0)) - \theta_B^0 \tau.} 
In the classical field theory, in such a
configuration, at energies below the scale $|z_0|/2\pi \alpha'$ we have
the $SU(N)\times SU(N+M)$ gauge theory with two bifundamental
hypermultiplets\foot{There is also a decoupled $SU(M)$ $\cn=2$ SYM theory
coming from the wrapped D5-branes at $z=-z_0$.}. 
In the limit of $z_0 \to \infty$, keeping $z$ constant, we get simply
\eqn\thetatwo{\theta = -2iM\ln(z/z_0) - \theta_B^0 \tau,} as above.
However, to be able to take $z_0
\to \infty$ we have to take $N \to \infty$ at the same time, if we
want to avoid the 5-form flux becoming negative (and the solution
becoming singular). Thus, the solutions may be identified with this large $N$
limit, and they do not make sense for finite $N$. As long as we do not take
$z_0 \to \infty$ and we remain with $\theta = -2iM\ln(z/(z+z_0)) -
\theta_B^0 \tau$, we can
have (as described in \newpol) consistent solutions also for finite
$N$, where the ratio between the radial coordinate $r_e$ where
the 5-form flux vanishes and $|z_0|$ behaves as $e^{-N\pi / g_s M}$ in the
large $N$ limit.

For simplicity, let us focus now on the solution $\theta =
-2iM\ln(z/z_0) - \theta_B^0 \tau$ 
(remembering that it should be interpreted via a large
$N$ limit as above). The supergravity equation for the 5-form field
has a source proportional to the wedge product of the two 3-forms in
the theory. This results in a D3-brane charge density in the solution,
localized completely at the orbifold singularity, and
proportional to $|\del_z \theta|^2 = 4M^2 / |z|^2$. The total 5-form flux
localized
between some circle of radius $r_1$ in the $z$-plane and a circle of
radius $r_2 > r_1$ in this plane is given by $M \ln(g_s M r_2 / \pi r_1)$.
When we reduce the radius by a factor of $e^{\pi / g_s M}$, the flux
is reduced by $M$, and
$\theta_B \to \theta_B + 2\pi$. Since $\theta_B$ is periodic with
period $2\pi$, the solution is thus self-similar under such shifts,
with an appropriate decrease of $N$ and appropriate shifts resulting
from the shift in $\theta$.

What is the interpretation of this decrease of $N$ ? At first sight
this seems to imply some sort of strange duality between the
$SU(N)\times SU(N+M)$ gauge theory and the $SU(N-M)\times SU(N)$
theory that we get after $N \to N-M$, as suggested in a similar
configuration in \klestr. However, we would like to suggest that the
interpretation in this case is much more mundane. We claim
that the D3-brane flux in the string theory can be
interpreted as corresponding to appropriate VEVs in the dual field
theory. Namely, we suggest that in the field theory the source for the
5-form can be thought of as coming not from 3-form fields (which have
no analog in the field theory), but rather from an actual distribution
of D3-branes and/or wrapped D5-branes
(corresponding to the field theory being at a specific
point in its moduli space). This means that
in the field theory there should be $M$ eigenvalues
of each of the two adjoint fields which correspond to positions in the
band between $r_1$ and $e^{\pi / g_s M} r_1$, for any $r_1$\foot{In the
supergravity limit we cannot say exactly where in the band these
eigenvalues lie because the width of the band is very small, so we
will leave this question open.}. Note that
we suggest that the eigenvalues of the two adjoint fields in this band
are equal, $z_j = \tilde{z}_j$, so they are not a source for the
2-form fields $\theta$. The full string theory
backgrounds of \refs{\newpol,\bdflmp} may then be identified with the
large $N$ limit of a configuration on the moduli space of the field
theory which has $M$ anti-D5-branes (= eigenvalues of $\varphi_2$)
at some $(-z_0)$ (which goes to infinity in the
large $N$ limit), $M$ eigenvalues of $\varphi_1$ and $\varphi_2$ in each band
between $r$ and $e^{\pi / g_s M} r$, and finally (when we ``run out of
eigenvalues'') $M$ D5-branes at some radius $r_e \simeq |z_0| e^{-N
\pi / g_s M^2}$ which constitute an enhan\c con. Depending on the
exact location of the eigenvalues, in each band the gauge group
$SU(N+M)\times SU(N)$ is spontaneously broken to some subgroup of
$SU(N)\times SU(N-M)\times U(M)\times U(M)$. The fields of the extra
$U(M)\times U(M)$ presumably correspond to some modes of the string theory
(living both off and on the orbifold) which tend not to propagate to
radii smaller than this band; we will discuss this further below. The
moduli of the extra $U(M)\times U(M)$ are not directly visible in the string
theory just like the original moduli were not visible, but we can make
them visible by changing the solution (``moving around the wrapped
5-branes'') as described above.

\subsec{Pros and Cons of the Higgsing Interpretation}

What is the evidence for this interpretation ? First, we claim
that it is the simplest possible interpretation of the background,
which does not require any unknown dualities, but just simply
translates the sources for the 5-form and 3-form fields to positions
in the moduli space.

Another test of the validity of this identification comes from
looking at the moduli space metric. The metric for a D3-brane probe,
corresponding to moving eigenvalues of both adjoint matrices together,
is flat and boring in the configuration we are discussing (since the
dilaton and axion are constant). However, the metric for a wrapped
D5-brane is more interesting, and was computed in
\bdflmp. Interestingly, this effective metric does not depend on the
function $Z$, which appears in the background metric and 5-form but
cancels out completely in the wrapped D5-brane metric. The result is
simply 
\eqn\metric{g_{ab} = \delta_{ab} {1\over 8\pi g_s} \theta_B = \delta_{ab}
{1\over 8\pi g_s} (\theta_B^0 + 2M g_s \ln|z/z_0|).} 
This is exactly
the same result we get in the field theory, identifying the D5-brane
position with an eigenvalue of $\varphi_1$,
if we identify (as discussed in \newpol) $4\pi / g_{SU(N+M)}^2 =
\theta_B^0 / 2\pi g_s$. The logarithm in the field theory arises at 1-loop in
the configuration described above, independently of where we put the
D3-branes (= equal adjoint eigenvalues) which do not affect the 1-loop
result. A non-renormalization theorem guarantees that there are no
additional perturbative corrections to the moduli space
metric. Additional non-perturbative corrections can arise from
instantons, but are negligible (in the large $N,M$ limit) as long as
the effective coupling does not diverge, namely, whenever \metric\ 
is non-zero.

However, the metric \metric\ does in fact vanish quite close to $z_0$,
when 
\eqn\zone{|z| = z_1 \equiv |z_0| e^{-\theta_B^0 / 2 M g_s},}
and it seems to become negative when $|z| < z_1$. How should we
interpret this ? 
From the supergravity point of view, this divergence of the effective
coupling is
very similar to the one which arises at an enhan\c con. Attempting to
move a wrapped D5-brane beyond this radius leads to a
non-supersymmetric configuration and thus costs energy. We suggest
that, as in the enhan\c con case, the field theory configuration space
does not allow the corresponding distribution of eigenvalues in which
a single eigenvalue is taken below $z_1$. Since we do not have an
explicit expression for the Seiberg-Witten curve in this case we
cannot check this directly, but it seems to be a prediction of the
AdS/CFT correspondence. We will call the circle $|z|=z_1$ a generalized
enhan\c con ring, and we will discuss the physics there in more detail
below. Note that this generalized enhan\c con ring is different from the
original enhan\c con in that it is not constituted of branes but
of supergravity fields (and some additional fields described in the 
next subsection), and the solution does not change significantly
as one goes through it. In particular, as
we discuss in a moment, other probes can go through this generalized
enhan\c con without feeling it.

Even though we cannot take a wrapped D5-brane to $|z| < z_1$, the
gravitational background there is not singular, and we can probe it
with other probes. For example, we can use a probe composed of $n$
D3-branes plus a single wrapped D5-brane (or, equivalently, $n+1$
wrapped D5-branes and $n$ wrapped anti-D5-branes). It is easy to
compute the metric for such a probe in the background of
\refs{\newpol,\bdflmp}, 
and we find 
\eqn\newmet{g_{ab} = \delta_{ab} {1\over 8\pi g_s} (2\pi n +
\theta_B^0 + 2M g_s \ln|z/z_0|)} 
(we can compute this directly, or just
note that this configuration may be derived from the previous one by
taking $\theta_B \to \theta_B + 2\pi n$; note that the latter fact
means that this is the natural probe for the theory at
$|z|\simeq |z_0| e^{-n\pi/g_s M}$). Thus, for this probe the
metric does not become singular until a different radial coordinate
$|z|
= |z_0| e^{-(2\pi n + \theta_B^0) / 2 M g_s}$. This is exactly the same
result that we get in the field theory for a computation of the
effective metric for shifting together $n+1$ eigenvalues of the
$SU(N+M)$-adjoint field $\varphi_1$ and $n$ eigenvalues of the
$SU(N)$-adjoint field $\varphi_2$, as expected. Similarly, for any
other probes, as long as their effective metric is non-singular, we
find exact agreement between the 1-loop field theory result and the
supergravity result.

Note that it is crucial for these agreements that the wrapped D5-brane
always corresponds to the bigger gauge group (i.e. that we always have
$M$ wrapped anti-D5-branes at infinity). If we had an interpretation
as in \klestr, in which the $SU(N+M)\times SU(N)$ group becomes an
$SU(N-M)\times SU(N)$ group with the wrapped-D5-brane gauge group
becoming $SU(N-M)$, the sign of the derivative of the effective metric 
on a wrapped-D5-brane
probe would change, and there is no sign of this in the supergravity
background.

The main evidence for our claim about the field theory VEVs
comes from a direct computation of the VEVs of the
corresponding operators in the string theory dual. The
sources of the bulk fields (in the untwisted sector), such as the
function $Z$ which appears in the metric and the 5-form, are (by
construction) exactly the
same as the sources that would arise from a distribution of D3-branes
of the form described above \bdflmp. Thus, if we construct the theory from an
$SU(N+M)\times SU(N+M)$ theory as described above, we can compute the
VEVs in the field theory by the usual methods of the AdS/CFT
correspondence (since the background in this case is asymptotically
AdS), and the VEVs will be consistent with our interpretation here,
and not with an interpretation in which the VEVs vanish and there is
some duality relating the theories at different scales. Note that to
distinguish the two possibilities we have to use untwisted sector
fields, which correspond to operators like
$\tr(|\varphi_1|^{2n})+\tr(|\varphi_2|^{2n})$ (this operator is identified with
some combination of the metric and 5-form fields), 
rather than twisted sector fields which
correspond to operators like $\tr(|\varphi_1|^{2n})-\tr(|\varphi_2|^{2n})$. 
The
latter operators do not depend on how we distribute the D3-branes so
they cannot distinguish the two possibilities. Note also that even
though the solution of \refs{\newpol,\bdflmp} is only invariant under
an $SU(2)\times SU(2)$ symmetry rotating the directions $x_6-x_9$, the
source for $Z$ is invariant under an additional $U(1)_R$ symmetry changing
the phase of $z$. Thus, only operators which are singlets of $SU(2)\times
SU(2)\times U(1)_R$ will obtain VEVs in the vacuum described by this solution.
The simplest such operator is the orbifold generalization of the operator
$\tr(2\Phi_4^2 + 2\Phi_5^2 - \Phi_6^2 - \Phi_7^2 - \Phi_8^2 - \Phi_9^2)$
in the $\cn=4$ theory, where $\Phi_i$ is the scalar field whose VEV is
related to the position in $x_i$. This operator is of the form
$[2\tr(|\varphi_1|^2) + 2\tr(|\varphi_2|^2) + {\rm hypermultiplet\ fields}]$. 
These operators are not chiral in the
$\cn=2$ theory, but they are still present in the supergravity and their
VEVs are non-zero, confirming the Higgsing interpretation.

There is one problem with our interpretation, which is the main
motivation for believing that a
more complicated interpretation involving some strong-coupling dual
might be required. The problem is that the 1-loop running of the two
gauge couplings is constant in the vacuum we are considering, below
the scale $|z_0|/2\pi \alpha'$ where we put the $M$ anti-D5-branes. The $SU(N)$
coupling becomes weaker as we go down in energy, but the $SU(N+M)$
coupling becomes stronger, and (using the 1-loop result) diverges at a
scale quite close to $|z_0|/2\pi \alpha'$ 
(in fact, it diverges exactly at the
scale $z_1/2\pi \alpha'$, with $z_1$ the position
where the wrapped D5-brane kinetic terms vanish as discussed
above). One possible way to interpret this is that when this coupling
becomes infinite, we have to make some strong-weak coupling duality
and go over to some other gauge theory, as suggested for a similar
situation in \klestr. However, there is no known dual for the gauge
theory we are discussing, and certainly the $SU(N+M)\times SU(N)$ and
$SU(N-M)\times SU(N)$ $\cn=2$ theories are not equivalent in any sense
(for instance, they have different dimensions for their moduli space). Our
suggestion is that the gauge coupling indeed becomes large at the position
of the first generalized enhan\c con ring, and presumably non-perturbative
corrections to the beta function become important so we do not really
know what happens to the running coupling 
below this scale. However, it seems that the
running coupling as a function of energy is not directly related
to anything we measure in our background; we can only
directly relate measurements of $\theta(z)$ 
to the effective moduli space metric,
as described above. Thus, it does not seem logically inconsistent to
suggest that the running coupling is strong, though the effective coupling on
some components of the moduli space is still (relatively) weak, as described
above. In any case, even if a strong-weak coupling duality is required
to understand this issue, we claim that it is not related to the
reduction in the size of the gauge group, which seems to arise from a Higgs
mechanism as described above.

Another apparent problem with our interpretation is that it violates the usual
UV/IR matching of radial positions with energy scales in the field
theory (this is also related to the previous problem). We suggest that the
moduli of the theory are spread out over different radial positions,
even though they are all low-energy fields in the field theory. Our
justification for this is that the UV/IR correspondence is only
understood in backgrounds which are asymptotically AdS, and even there
the matching can be quite subtle (as in the solutions of
\polstr\foot{In solutions of \polstr\ involving more than one 5-brane,
the massless particles of the field theory live on the 5-branes which are
at finite radial positions, while the string theory fields at small
radii correspond to massive particles of the field theory.}), so
we do not see this as a concrete problem. Note in particular that our
conjecture implies that putting in a cutoff at some radial position 
in the string theory is not simply equivalent to an energy cutoff in
the field theory, but again there are already various examples of this.

\subsec{Comments on the Generalized Enhan\c con Rings}

Finally, let us try to discuss in more detail the physics of the
generalized enhan\c con rings. 
The generalized enhan\c con rings occur when the kinetic term
of a probe wrapped D5-brane attached to some number of D3-branes
vanishes; this means that $\theta_B$ is an integer multiple of
$2\pi$. At first sight this suggests that we have tensionless strings
on the generalized enhan\c con ring, but we should be more careful because
$\theta_C$ does not generally vanish (in the sense of being $2\pi$
times an integer) there. In fact, in the background \thetatwo\ 
$\theta_C$ vanishes only on $2M$
(equally-spaced) ``enhan\c con points'' 
on the generalized enhan\c con ring, where $\theta=2\pi
(n-m\tau)$ for integer $n,m$. 
So, we only get light fields beyond the supergravity modes (and the
twisted sector tensor multiplet)
at these points. Note that a probe NS 5-brane (attached to some number
of D3-branes) would have a vanishing kinetic term when $\theta_C$ is
an integer multiple of $2\pi$, so at these points the kinetic terms
for both probes vanish. However, we do not have a direct field theory
interpretation for such a probe (presumably it is related to
condensing some ``magnetic'' degrees of freedom in the field
theory). The low-energy field theory at the ``enhan\c con points'' is not
clear. It seems that it is not a free field theory, so it is not clear
to what extent the fields living at the enhan\c con point can be
identified with the $2M$ ``extra'' moduli corresponding to breaking
$SU(N+M)\times SU(N)\to SU(N)\times SU(N-M)\times G$ (where $G$ is a
subgroup of $U(M)\times U(M)$ of rank $2M$). By small changes in
wrapped D5-brane positions we can move the ``enhan\c con points'' around,
so it seems that they are related to the moduli, but the precise
relation seems to be quite complicated.

The low-energy field theory living on the orbifold
fixed line generically involves a free tensor multiplet, three of whose
scalars correspond to blowing up the orbifold while the other two
scalars are periodic and correspond to the 2-form fields integrated over
the vanishing 2-cycle. For a particular value of the two periodic
scalars we have additional light degrees of freedom. Note that this is
not the same as the low-energy theory of the ``little string theory'',
in which only one scalar is periodic; the ``little string theory''
arises in the limit $g_s \to 0$ of the $\IR^4/\IZ_2$ singularity,
in which the periodicity of $\theta_B$
diverges (in physical units). In the background we are discussing it
is clear that this periodicity is important, so one cannot discuss it
in terms of the decoupled theory living on the $\IR^4/\IZ_2$
singularity. If we look at a particular generalized 
enhan\c con ring, $\theta_B$
is constant along it while $\theta_C$ varies as $2M$ times the
angle. From the point of view of the low-energy field theory on the
fixed line, one of
the periodic scalar fields is linear in the angle. In this case it seems that
one can discuss the physics in terms of the physics of ``little string
theories'', but the behavior of
``little string theories'' in configurations of this type is not known.

\newsec{Theories Related to Branes on a Conifold Singularity}

The case of branes on a conifold was extensively discussed in
\refs{\klewit,\gubser,\klegub,\klenek,\kletse,\klestr} 
so we will be relatively brief in reviewing it
here. The low-energy theory for $N$ D3-branes on a conifold is
believed to be the low-energy limit of an $SU(N)\times SU(N)$ $\cn=1$
gauge theory with 2 pairs of bifundamental chiral multiplets,
$A_1,A_2$ in the $(\bf{N,\bar{N}})$ representation and $B_1,B_2$ in
the $(\bf{{\bar N},N})$ representation, and with a quartic
superpotential proportional to 
\eqn\suppot{W = \tr(A_1 B_1 A_2 B_2 - A_1 B_2 A_2 B_1).} 
This theory flows to a superconformal theory (in fact, a fixed
line of superconformal theories \leistr) in the IR. This SCFT
is believed \klewit\ to be dual to type IIB string theory on the
near-horizon limit of the D3-branes on the conifold, which is
$AdS_5\times T^{1,1}$. In particular, the classical moduli space of this theory
is exactly $N$ copies of the conifold (divided by $S_N$), and may be
identified in terms of configurations with different distributions of
D3-branes as described above. The classical theory has an $SU(2)\times
SU(2)\times U(1)_B\times U(1)_R$ global symmetry, where the two $SU(2)$
factors rotate the fields $A_i$ and $B_i$, respectively, while under
$U(1)_B$ the fields $A_i$ have positive charge and the fields $B_i$
have negative charge. All these symmetries are unbroken also in the
quantum theory (at the origin of moduli space). The $SU(2)\times
SU(2)\times U(1)_R$ symmetry is identified with the geometrical
symmetry of $T^{1,1}$, while $U(1)_B$ is identified with the gauge
symmetry coming from the 4-form field of type IIB string theory
integrated over the 3-cycle in $T^{1,1}$.

If we add to the $N$ D3-branes
$M$ additional D5-branes wrapped on the 2-cycle which
vanishes at the conifold point, it seems that we
get an $SU(N+M)\times SU(N)$ theory with the same field content and
superpotential. As above, this theory is not asymptotically-free
(although both gauge couplings are asymptotically-free at 1-loop), so
it is not clear how to define it as a field theory. In
\refs{\kletse,\klestr} it was
shown that the near-horizon limit of this background leads to a
configuration very similar to the one described in the previous
section, in which the 5-form flux
grows logarithmically as we go to large radii, and there is a
scalar field theta
(coming from the integrals of the 2-forms over the 2-cycle)
which also varies. Unlike the $\cn=2$
case, in this case it is not known how to obtain this configuration
from a well-defined field theory in some limit, though presumably the
background is still in some sense dual to the large $N$ limit of this
$SU(N+M)\times SU(N)$ theory. Another important difference from the
$\cn=2$ case is that in the solution of \klestr\ 
the volume of the 2-cycle does not
vanish (except at the minimal radial position), and there is no flat
direction corresponding to moving around wrapped 5-branes. This
corresponds to the fact that the classical
moduli space of the $SU(N+M)\times SU(N)$ theory is
still just $N$ copies of the conifold, and there are no branches
corresponding to ``fractional branes''.

In the quantum theory with $M > 0$, the $U(1)_R$ symmetry is
anomalous, and the superpotential receives quantum corrections. This
leads to a correction in the quantum moduli space, and it was argued
in \klestr\ that some branches of the moduli space 
look like D3-branes moving on the deformed conifold
rather than on the conifold. This agrees with the string theory dual
found in
\klestr, which can indeed include any number of D3-branes moving on
the deformed conifold (at least when their back-reaction on the
background can be neglected). The deformed conifold no longer has the
$U(1)_R$ symmetry, but it still has the $SU(2)\times SU(2)$
symmetry. When
$p \equiv N {\rm \ mod\ } M$ vanishes, the string theory dual has
a good supergravity approximation, with no additional branes. For
non-zero $p$ the solution of \klestr\ involves also (at least) $p$
D3-branes moving in the background. These D3-branes break the
$SU(2)\times SU(2)$ symmetry. In the field theory, the corresponding VEVs 
can be thought of as breaking
the $SU(N+M)\times SU(N)$ group to $SU(N+M-p)\times SU(N-p)$, which is
in the class of theories with $p=0$. Thus, it seems to be sufficient to
understand the behavior in the case of $p=0$, and we will focus on
this case from here on.

It was suggested in \klestr\ that the interpretation of the decrease
in the value of $N$ in this case should be via Seiberg
duality \seiberg. As in the previous section, at 1-loop the gauge
coupling of the $SU(N)$ factor becomes weak as we go down in energy,
while that of the
$SU(N+M)$ factor becomes strong and, if we define it to be constant at
some UV cutoff, diverges quite close to the cutoff. 
One can argue \refs{\klenek,\klestr} 
that the exact renormalization group flow indeed
causes the coupling of the $SU(N+M)$ gauge group to become very strong,
and the authors of \klestr\ 
argued that this is the same as the behavior of this
theory at low energies, where (if we can ignore the dynamics of the
$SU(N)$ gauge group) it is dual \seiberg\ 
to an $SU(N-M)$ gauge theory. Therefore, they
argued that the reduction in $N$ comes from a duality rather than a
Higgs mechanism, and they showed that after the duality one gets a
similar theory with gauge group $SU(N-M)\times SU(N)$ instead of the original
$SU(N+M)\times SU(N)$, in agreement with the decrease in the 5-form flux.

The quantum-corrected moduli space of these theories seems to be very
complicated, and to contain many different types of branches\foot{
The rest of this section is based on discussions
with I. Klebanov and M. Strassler. I am grateful to I. Klebanov and
M. Strassler for correcting some mistakes in an earlier version of
this section.}. The
$SU(N+M)\times SU(N)$ theory seems to have a branch of (complex)
dimension $3N$, which looks like $N$ D3-branes moving on the deformed
conifold (and which would arise from the near-horizon limit of a
configuration where the D3-branes are taken slightly off the conifold
point). Describing such a branch in the dual theories with a smaller
value of $N$ is apparently quite complicated, and requires
including also some massive fields in the dual theories. It was
suggested in \klestr\ that the quantum-corrected moduli space of the
$SU(N+M)\times SU(N)$ theory includes $M$ branches of dimension
$3(N-kM)$ for every $k=0,1,2,\cdots,[N/M]$.  It was shown in \klestr\ that
the moduli space indeed had this form for $N=M$; it would be
interesting to verify that this is the form of the moduli space also
for $N > M$. In the particular case of $p=0$ the smallest branches in
fact have dimension one rather than zero\foot{We thank M. Strassler for
notifying us of this correction to the claims of \klestr.}, 
and involve VEVs of ``baryonic
operators'' (which we will describe below).
These branches are
part of the moduli space of all the $SU(N+M)\times SU(N)$ theories, for
any $N=kM$, and can be described just in terms of the massless fields
in all of these theories. 

To understand the interpretation of the decrease in $N$ we need to
understand which branch in the moduli space the background of \klestr\
corresponds to. The Higgsing interpretation makes sense if the theory
is in the branch with maximal dimension (the dimension is infinite in
the large $N$ limit just like in the case of the previous section),
which can be directly related to some distribution of branes, as in
the previous section. We will
call this the ``mesonic branch'' since the positions of D3-branes are
generally related to eigenvalues of the ``meson matrices'' $N_{ij}
\equiv A_i B_j$, which are in the adjoint of $SU(N)$. On the
other hand, the duality interpretation seems to be most useful if the
theory is on the branch of lowest dimension, which can be described
solely in terms of the massless fields in all the different dual
theories. We will call this branch the ``baryonic branch'' since the
computations of \klestr\ suggest that the VEVs of meson operators vanish in
this vacuum, while baryonic operators obtain VEVs.

How can we distinguish which branch are we in ? We can try to compute
VEVs of gauge-invariant operators in the background, but it is not
clear exactly how to do this in this case (unlike the previous
section), since we do not know how to obtain this theory from a limit
of theories which are under control. So, all we can do is
try to use general arguments to constrain the
VEVs. Since the background of \klestr\ is invariant under
$SU(2)\times SU(2)$, it is clear that only VEVs of operators invariant
under this symmetry can be non-zero. VEVs for eigenvalues of the mesonic
operators $N_{ij}$ break this symmetry, so it seems that we are not in
a mesonic branch, but one could get around this by saying that in the
large $M$ limit the D3-branes are smeared in an $SU(2)\times
SU(2)$-invariant way (as in other backgrounds describing
configurations on the moduli space, like the ones described in
\fgpw). In such a distribution only $SU(2)\times SU(2)$-invariant
operators of the form $\tr(\prod_l A_{i_l} B_{j_l})$ would obtain
VEVs. Unfortunately, one can show 
that all of these operators are non-chiral and
do not appear in the supergravity spectrum, so
it is hard to compute these VEVs (both in the
field theory and in the supergravity). However, it seems that in a
configuration like this, even in the large $N$ limit the $SU(2)\times
SU(2)$ symmetry would be spontaneously broken, so there should be
Goldstone bosons in the background, while the background of \klestr\
seems to have a mass gap\foot{Note that the backgrounds of \fgpw\ also had
a mass gap in the supergravity approximation, but this approximation
breaks down there and the background includes additional branes which
could carry the Goldstone bosons. On the other hand, supergravity seems to be
a reliable approximation 
in the background of \klestr.}. Thus, it seems more natural to
identify this background with the ``baryonic branch'' of the field
theory, which preserves the
$SU(2)\times SU(2)$ symmetry.

In addition to the ``mesonic operators'' of the form $\tr(\prod_l
N_{{i_l} {j_l}})$,
the gauge theories with $p=0$ contain also ``baryonic operators'' of
the form 
\eqn\baryons{{\cal B} \equiv [(A_i)^N]^{(N+M)/M}; \quad {\overline {\cal B}}
\equiv [(B_j)^N]^{(N+M)/M},} 
where in $(A_i)^N$ we contract the indices to form an $SU(N)$-singlet
in the $M$'th antisymmetric representation of $SU(N+M)$, and then we
contract the $SU(N+M)$ indices using an epsilon symbol (or we could do
the contractions in the opposite order with the same
result). All the operators of this type have a non-zero
$U(1)_B$ charge, and they can have various $SU(2)\times SU(2)$ quantum
numbers depending on how we choose the $SU(2)$ indices $i$. These
operators include an $SU(2)\times SU(2)$-singlet which could acquire a
VEV in the vacuum corresponding to the solution of \klestr, and in
fact it should obtain such a VEV according to the field theory
analysis of the ``baryonic branch'' in \klestr\ (at least for $N=M$). 
Such a VEV should spontaneously break the $U(1)_B$
symmetry. This $U(1)_B$ symmetry is not visible in the background of
\klestr, which is consistent with this claim. 

The operators $\cal B$
may be identified (by a generalization of the analysis of \klegub), in the
region where the 5-form flux is $N+M$, with configurations consisting of
$(N+M)/M$ D3-branes wrapped on the 3-cycle, 
with $M$ strings ending on each one, and a single
D5-brane wrapped on the compact 5-cycle, 
with the other ends of the $N+M$ strings ending on the
D5-brane\foot{The number of strings ending on each brane is determined by
charge conservation after taking into account the RR fields in the
background.}. When one moves such a ``baryon particle'' in the radial
direction such that the 5-form flux decreases by $M$
and $\theta_B$ shifts by $2\pi$,
the D5-brane acquires $(-1)$ units of
D3-brane charge. Then, it can separate into a D5-brane 
with no D3-brane charge and an
anti-D3-brane, and the anti-D3-brane
can annihilate one of the D3-branes. This exactly reproduces
the change in the baryon operator as one decreases $N$ (and it is
consistent with the behavior of this operator under Seiberg
dualities). As we decrease the radial position to its minimal value
(where the 5-form flux vanishes and the 2-cycle contracts to zero
size), all the D3-branes annihilate and one
is left with nothing. In the gauge theory this corresponds to the fact
that the baryon is a singlet of the final $SU(M)$ gauge group, and it
proves that indeed the baryon does not carry any conserved charge in
the background of \klestr, so the $U(1)_B$ is indeed broken. Since the
$U(1)_B$ is spontaneously broken in the field theory, 
there should be a corresponding
Goldstone boson in the background, and it would be interesting to
identify this state. Note that unlike the $SU(2)\times SU(2)$
Goldstone bosons, this state should couple only to non-perturbative
states like the one described above, so it is more complicated to
identify it, and it would not lead to massless poles in the scattering
of states corresponding to supergravity fields.

Thus, it seems that the background of \klestr\ corresponds to the
``baryonic branch'' of the field theory\foot{Note 
that Seiberg duality applies also to baryonic
branches of the moduli space when all quantum corrections are properly
taken into account \oldme.}. Therefore,
the decrease in $N$ in this background seems to correspond to
a duality interpretation rather than to a Higgsing
interpretation\foot{A relation between Seiberg duality and Higgsing
in a ``baryonic branch'' appeared in the work of \aps\ on
mass-deformed $\cn=2$ gauge theories. It would be interesting to
understand if there is any relation between those results and our
results here; there could perhaps be a direct connection since naively
when we add mass terms to the $\cn=2$ theory described in the
previous section we get the theory described in this section.}.
Note that this identification means that configurations
with different numbers of D3-branes moving in the background of
\klestr\ are generally not on the same branch of the moduli space.

If our identification is correct,
our discussion suggests that
there should also be generalizations of the background of \klestr\
which would correspond to other branches of the moduli space,
including branches where a
Higgs interpretation would be more appropriate. Another solution
with the same asymptotic behavior was found in \leoark. This solution
involves a resolved conifold rather than the deformed conifold which
appears in \klestr.
The solution of \leoark\ is singular and has a repulson in it, but it might
be possible to resolve this singularity by the enhan\c con mechanism
and replace it with some distribution of branes which is a consistent
background of string theory. The solution of \leoark\ was claimed in
\cglp\ not to be supersymmetric, based on the behavior of the 3-form
fields in that solution. However, this could change once the
singularity is resolved, since we expect the enhan\c con to be a
source for the 3-form fields and to change the metric
(presumably the enhan\c con consists of
D5-branes wrapped on the 2-cycle). So, we will assume here that there is a
resolution of the singularity (or a generalization of the solution of
\leoark) which is supersymmetric. In such a case this background should be
a different vacuum of the field theory we discussed in this section.
We conjecture that this background may correspond 
to a point on the ``mesonic
branch'' of the moduli space, where there is a distribution of the
eigenvalues of $N_{ij}$ which matches the source for the 5-form field
in the solution. Again, we do not know how to verify this conjecture
directly, but it seems likely that this should be the case, since
already in the conformal case \klewitt\ resolving the conifold was
interpreted as giving VEVs to the mesonic operators. Note that,
assuming that the resolution of the singularity in 
this background involves an enhan\c con,
it may not have a mass gap, which would be consistent with our expectation
for having $SU(2)\times SU(2)$ Goldstone bosons in the background
corresponding to the ``mesonic branch''. As additional evidence we
note that in the resolved conifold the 2-cycle never collapses, so we
may not be able to annihilate to nothing the baryon particle described above,
suggesting that perhaps in this background the $U(1)_B$ symmetry is
unbroken. All this is very weak evidence
for our conjecture; it would be interesting to perform more tests of 
this identification.

We should emphasize that whenever we are talking of an interpretation
for the decrease in $N$, we are really talking about what happens in
the weakly coupled theory, since only there we can really see that we
have some particular gauge group, and what happens to it when we
change the energy scale. The supergravity backgrounds all correspond
to strong coupling (large $g_s M$), so the more precise statement of
our suggestion is that if we
continue these backgrounds to weak coupling, then the solution of \klestr\
would have an interpretation via Seiberg duality, while that of
\leoark\ might have an interpretation in terms of a Higgs
mechanism. It is not clear what these interpretations mean directly in
the strongly coupled theory. Naively the change in the theta angle can
be interpreted as a running coupling, which was interpreted in
\klestr\ as supporting the duality interpretation, but we saw in the
previous section that such an interpretation does not always seem to
work.

It would be very interesting to
understand better the exact quantum superpotential in these theories,
in order to verify the claims of \klestr\ about the quantum moduli
space. We gave evidence for identifying particular branches in this
moduli space with particular string theory solutions, but our evidence
is far from conclusive, and it would be nice to substantiate these
identifications. In particular,
it would be interesting to close the possible loopholes in
our identification of the configuration of \klestr\ with the
``baryonic branch'' in the moduli space, for example by identifying
properly the $U(1)_B$ Goldstone boson (which presumably comes from the
RR fields in the background of \klestr), and to test our suggestion
that the configuration of \leoark\ could correspond to the ``mesonic
branch'' of the same theory. Of course, this requires a resolution of
the singularity in the solution of \leoark. Perhaps there are
different ways of resolving this singularity, that would correspond
to different branches in the field theory moduli space, or there could
be additional solutions (perhaps involving generalizations of 
the deformed conifold and resolved conifold metrics)
that would correspond to the other branches. Alternatively, it is
possible that there is no simple string theory dual for the ``mesonic
branches'' of the moduli space.

Hopefully, the results we presented here will enable a better
understanding of holography in backgrounds with varying flux. In
particular, it would be nice to have a better understanding of exactly
how to define the corresponding large $N$ field theories directly in
field theory terms. We saw that such an understanding exists in the
$\cn=2$ case, but it does not yet exist in the $\cn=1$ case as far as
we know.

\vskip 1cm
{\bf Acknowledgments:}
I would like to thank M. Berkooz, S. Elitzur, A. Giveon, N. Itzhaki, 
C. Johnson, B. Kol,
E. Rabinovici, A. Schwimmer, J. Simon and J. Sonnenschein
for useful discussions, and
especially I. Klebanov, J. Polchinski and M. Strassler for many useful
discussions and correspondence. I thank the Aspen Center for Physics
for hospitality when I first started thinking about these issues.
This work was supported in part by the IRF Centers of Excellence
Program, by the European RTN network HPRN-CT-2000-00122, by
the United States-Israel Binational Science Foundation (BSF), and by Minerva.

\listrefs

\end